%% file: main.tex
\documentclass{article} 

\AtBeginDocument{%
  \providecommand\BibTeX{{%
    \normalfont B\kern-0.5em{\scshape i\kern-0.25em b}\kern-0.8em\TeX}}}

\usepackage{multirow} 
\usepackage{adjustbox} 
\usepackage{caption} 
\usepackage{colortbl} 
\usepackage{subfig} 
\usepackage{tablefootnote} 
\usepackage{algorithm2e} 
\RestyleAlgo{ruled}
\usepackage{xcolor} 
\usepackage{parnotes} 
\usepackage{tablefootnote} 
\usepackage{url} 
\usepackage{authblk}
\usepackage{hyperref} 

\usepackage{booktabs}

\newcommand\hc{ \rowcolor{orange!40}}

\newlength{\defbaselineskip}
\SetKwComment{Comment}{/* }{ */}

\begin{document}

\title{DQRM: Deep Quantized Recommendation Models} 

\author[1]{Yang Zhou} 
\author[2]{Zhen Dong} 
\author[3]{Ellick Chan} 
\author[3]{Dhiraj Kalamkar}
\author[5]{Diana Marculescu} 
\author[2]{Kurt Keutzer} 

\affil[1]{Carnegie Mellon University\\ \texttt{yangzho6@andrew.cmu.edu}} 
\affil[2]{University of California, Berkeley\\ \texttt{\{zhendong, keutzer\}@berkeley.edu}} 
\affil[3]{Intel\\ \texttt{\{ellick.chan, dhiraj.d.kalamkar\}@intel.com}} 
\affil[5]{University of Texas at Austin\\ \texttt{dianam@utexas.edu}} 

\date{} 

\maketitle

\input{s0_abstract.tex} 

\maketitle

\input{s1_introduction.tex}

\vspace{-1em} 
\input{s5_previous_works} 
\input{s2_methodology}

\input{s3_experiments}
\input{s6_conclusion}

\bibliographystyle{ACM-Reference-Format}
\bibliography{sample-base} 

\clearpage
\appendix

\input{s7_supplements}

\end{document}

%% file: s0_abstract.tex
\begin{abstract}
\normalsize
Large-scale recommendation models are currently the dominant workload for many large Internet companies. These recommenders are characterized by massive embedding tables that are sparsely accessed by the index for user and item features. The size of these 1TB+ tables imposes a severe memory bottleneck for the training and inference of recommendation models.
In this work, we propose a novel recommendation framework that is small, powerful, and efficient to run and train, based on the state-of-the-art Deep Learning Recommendation Model (DLRM). The proposed framework makes inference more efficient on the cloud servers, explores the possibility of deploying powerful recommenders on smaller edge devices, and optimizes the workload of the communication overhead in distributed training under the data parallelism settings. Specifically, we show that quantization-aware training (QAT) can impose a strong regularization effect to mitigate the severe overfitting issues suffered by DLRMs. Consequently, we achieved INT4 quantization of DLRM models without any accuracy drop. We further propose two techniques that improve and accelerate the conventional QAT workload specifically for the embedding tables in the recommendation models. 
Furthermore, to achieve efficient training, we quantize the gradients of the embedding tables into INT8 on top of the well-supported specified sparsification. We show that combining gradient sparsification and quantization together significantly reduces the amount of communication. Briefly, DQRM models with INT4 can achieve 79.07\% accuracy on Kaggle with 0.27 GB model size, and 81.21\% accuracy on the Terabyte dataset with 1.57 GB, which even outperform FP32 DLRMs that have much larger model sizes (2.16 GB on Kaggle and 12.58 on Terabyte). We open-sourced our implementation in 
\href{https://github.com/YangZhou08/Deep_Quantized_Recommendation_Model_DQRM}{DQRM code}.  
\end{abstract} 

%% file: s1_introduction.tex
\section{Introduction}
With the widespread adoption of Internet services, personalization becomes a critical function of the services provided by the current Internet giants. Every user's unique taste and preferences need to be catered to. With billions of internet users today, the need for recommendation models is more crucial than ever. According to ~\cite{ArchImpl19}, over 79\% of the entire Meta's cloud ML inference cycles are spent on the inference of various sizes of recommendation models. By Amdahl's law, a slight boost in recommendation model efficiency can yield a massive performance boost. On the other hand, if part of the inference workload can be migrated to edge devices, Internet service providers can save precious cloud resources, while the users can have less of their personal data sent to the cloud environment, which strengthens their personal data privacy. To achieve less costly recommenders, in this work, we propose a deep quantized recommendation model (DQRM) framework, where the models are both more efficient on the cloud environment, and are possible to fit on edge devices. Moreover, to address the necessity for periodic retraining in recommenders, the proposed framework is optimized for model training in the cloud-distributed environment.

\begin{figure*}[t] 
\includegraphics[width=\textwidth]{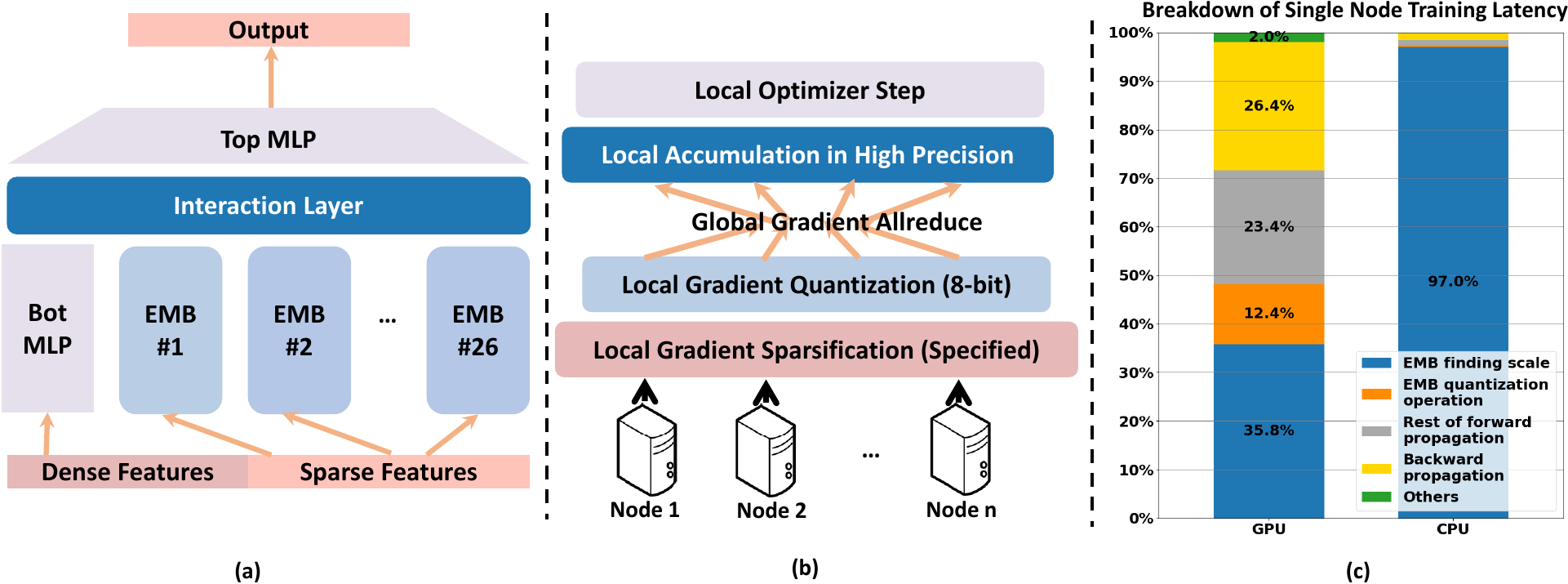} 
\caption{(a) shows the state-of-the-art large-scale recommendation model architecture. The model contains two types of layers: Embedding tables and MLP layers. (b) Our framework builds on top of specified sparsity and adds quantization to achieve additional gradient compression ratio. (c) shows a breakdown of the single-machine training time running DQRM in INT4; the majority of the training time on the GPU platform (left) is spent on finding the quantization scales, and even more so on the CPU node (right).} 
\label{fig:showing_and_breakdown} 
\end{figure*}

Specifically, designing deep learning-based recommendation models is challenging, because of the necessity to both process dense and sparse inputs. Successful previous works~\cite{DLRM19, wang2017deep} utilize massive embedding tables, each corresponding to a sparse feature category. Although embedding tables are proven to learn sparse features well, they are usually massive, and easily take up many GB, and in some cases even TB in DRAM memory. The massive model size makes memory the prime bottleneck in both training and inference. This severe problem motivates a huge effort in shrinking recommendation model size ~\cite{ginart2021mixed, shi2020compositional, kang2020learning, ko2021mascot, guan2019post, deng2021low}.
In our work, we base our framework on the Deep Learning Recommendation Model (DLRM) ~\cite{DLRM19}, which is the state-of-the-art for Click-Through-Rate (CTR) (~\cite{zhu2020fuxictr}) prediction (as in Figure ~\ref{fig:showing_and_breakdown} (a)), and we apply quantization and sparsification to enhance its efficiency.

Based on experiments and analyses, we find that DLRM models suffer from severe overfitting issues, with testing accuracy dropping steeply after the first one or two epochs of training. We propose to heavily quantize the DLRM model into ultra-low INT4 precision, and we show that QAT can greatly reduce the memory and computational costs while imposing a strong regularization effect to benefit DLRM training. 
As a result, we achieve comparable or even better testing accuracy than the full-precision models. Besides, our experiments indicate that conventional QAT can be highly inefficient in recommendation model quantization. To deal with this issue, we propose two techniques that reduce additional memory copies of large embedding tables during QAT and avoid the costly traversal of large tensors in memory to compute quantization scales. Our proposed methods significantly reduce time costs.

Besides the model size and inference costs, the training of recommendation models is also crucial in real-world applications. Recommenders are usually trained in highly distributed cloud environments, where the training time is dominated by heavy inter and intra-node communications overhead. In this work, we propose to apply communication sparsification and quantization jointly to shrink the workload. 
We show that using well-supported specified sparsification 
of gradients can compress the communication by three orders of magnitude. On top of that, we further quantize the communication from the massive embedding tables into INT8, shrinking the remaining size of communication by one-half during training, with a negligible accuracy degradation on the large Criteo Terabyte dataset. 
We show the backpropagation of the proposed framework in Figure ~\ref{fig:showing_and_breakdown} (b). 
Our contributions are summarized as follows: 

\begin{itemize}
\item Section \ref{methodology}, we propose to apply ultra-low precision quantization to alleviate the overfitting of DLRMs meanwhile achieving smaller model sizes; 

\item Section \ref{reduce_copy} and \ref{technique1}, we introduce two novel techniques to improve the speed of QAT on recommendation systems; 

\item Section \ref{met_distributed}, we enable more efficient training by jointly applying sparsification and quantization on communications; 

\item Section \ref{experiment_res}, our DQRM models with INT4 achieve an 8$\times$ reduction in model size, while obtaining 79.071\% accuracy (0.148\% higher than DLRM FP32 baseline) on the Kaggle dataset, and 81.210\% accuracy (0.045\% higher than DLRM FP32 baseline) on the Terabyte dataset. 

\end{itemize}

%% file: s5_previous_works.tex
\section{Previous Works} \label{pre_works} 

\textbf{Compressing Large-scale Recommendation Models} - DLRM ~\cite{DLRM19} is a typical and highly popular click-through-rate recommendation model designed and vastly deployed by Meta. 
Motivated by its importance, many previous works (~\cite{ginart2021mixed}, ~\cite{shi2020compositional}, ~\cite{yin2021tt}, ~\cite{desai2021random}, ~\cite{wu2020saec}) focus on compressing DLRM's model size. Since over 99\% of the DLRM model size is occupied by the embedding tables, previous efforts focus on shrinking the embedding tables without lowering weight precision. 
Our work focuses on neural network quantization, which is orthogonal and complementary to these works. \\ 
\indent Previously, quantization ~\cite{quant_survey} has been extensively studied on CNNs ~\cite{choi2018pact, esser2019learned, ultra_low_precision_wrap, degree-quant, yao2021hawq, xiao2023csq} and Transformers ~\cite{qbert, fan2019reducing, Zadeh_2020, liu2023noisyquant, kim2023squeezellm} and successfully applied to the Matrix Factorization and Neural Collaborative Filtering models \cite{kang2020learning, ko2021mascot}. 
Recently, some works have extended quantization to DLRM models but mainly focus on Post Training Quantization (PTQ). ~\cite{guan2019post} uses codebook quantization to quantize the embedding tables of DLRM models into INT4, while ~\cite{deng2021low} further quantizes the whole model into 4-bit. 
Both works revealed that PTQ introduces accuracy degradation, which motivates other quantization methods like Quantization-aware Training (QAT) to improve. Besides, low-precision training (LPT) of DLRM ~\cite{zhang2018training, xu2021agile, li2022adaptive} received attention. ~\cite{zhang2018training} and ~\cite{xu2021agile} train CTR models using quantized FP16 weights, while ~\cite{li2022adaptive} trains DCN ~\cite{wang2017deep} using INT8 weights during low-precision training and compress models into 2-bit and 4-bit. 
Different from the LPT works that achieve savings in training memory usage by allowing accuracy drop, our work explores QAT on DLRM models and aims for ultra-low precision during inference without accuracy loss, while making QAT much more efficient for large-scale recommendation models specifically. \\ 

\noindent \textbf{Efficient Training of Recommendation Models} - 
Training of large-scale CTR models is distributed in the real world, and communication bottlenecks the training time. 
Many previous works compress DLRM communication loads to speedup training time losslessly (~\cite{pumma2021semantic}) or lossily (~\cite{gupta2021training, yang2020training}) for a higher compression ratio. 
~\cite{gupta2021training} improves conventional Top-k sparsification on communication during DLRM's hybrid parallelism. ~\cite{yang2020training} quantizes the communication to ultra-low precision with tolerable accuracy drop through a novel error compensation scheme. Previously, ~\cite{sparcml} combines sparsification and quantization on distributed communication for CNN and ASR models. However, the merit of combining the two is not generalizable to different model architectures. In this work, we look into these techniques' effects on different parts of DQRM under the DP environment and combine specified sparsification and quantization together to further benefit DQRM training. 

%% file: s2_methodology.tex
\section{Methodology} \label{methodology} 

\begin{figure*}[t]
\includegraphics[width=\textwidth]{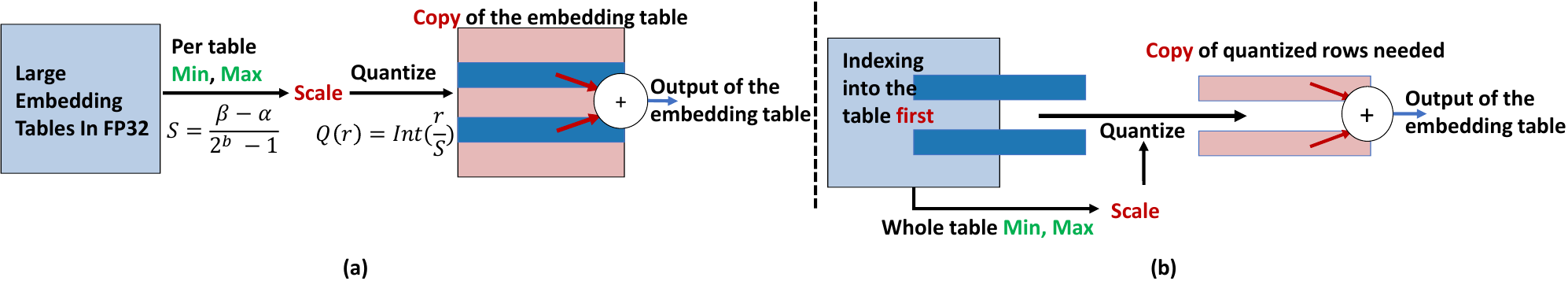} 
\caption{(a) Conventional QAT method, where the entire set of weights of the embeddings are copied and quantized. As embedding accesses are very sparse, this method is wasteful as it processes unused entries and worsens the memory bottleneck in QAT. (b) Our method to avoid the massive copy is by first performing the functional part of each table and then performing copying and quantization. In this way, only part of the weights that are actually used is copied and quantized, utilizing the memory more efficiently (Figure best viewed in color).}
\label{fig:conventionvsimproved} 
\end{figure*} 

\subsection{Reducing Memory of Unused Weights} \label{reduce_copy} 
DLRM-style recommenders have over 99\% of their model size occupied by embedding tables, unlike other popular architectures like CNNs or transformers.
Because of these giant embedding tables, previous works~\cite{ko2021mascot, ginart2021mixed, erdoganstochastic} have shown that training large-scale recommendation models is memory-bound instead of compute-bound. 
The memory bottleneck problem for DLRMs is further exacerbated if Quantization-aware training (QAT) is naively applied. 

As shown in Figure ~\ref{fig:conventionvsimproved} (a), in naive QAT, the entire weight tensor of convolutional or MLP layers is quantized since it is used in the subsequent computation. This requires two copies of the tensors to be stored: a low-precision and a full-precision shadow copy. However, as shown in Figure ~\ref{fig:conventionvsimproved} (b), 
since not all embedding weights contribute to subsequent computation, copying the entire embedding table in memory is unnecessary during quantization. 

Therefore, 
our solution is to run the embedding layer first and retrieve only the necessary sparse embedding vectors to quantize, so only used embedding vectors have low-precision copies stored. 
This approach ensures that only active embedding table entries are copied throughout the iteration, which is usually more than three magnitudes smaller than the total number of embedding vectors in the tables on average depending on the training batch size. 

\subsection{Periodic Update to Find Quantization Scale} 
\label{technique1} 
Naive QAT poses more challenges to DLRM quantization. 
During the quantization process, the first step is usually finding the quantization scale. 
Given the large embedding table size, this step is extremely costly as the program needs to traverse through an enormous chunk of contiguous memory. We run a breakdown of QAT training time shown in Figure ~\ref{fig:showing_and_breakdown} (c) under the single-node training setting for both GPU (training on Kaggle dataset) and CPU (training on Terabyte dataset). \footnotemark[1]
\footnotetext[1]{More experiment details can be found in Appendix \ref{additiondistributedtraining}.} 
For reference, without quantization, backward propagation workload dominates total single-node GPU training time: 5.8 ms out of 8.5 ms total. 
When QAT is applied, backprop (the yellow portion) no longer dominates, as forward propagation becomes significantly longer (indicated by the rest of the pillars). The primary reason for this increase in time is the operation of finding the scale (shown by the blue portion).
Finding the scale of large embedding tables occupies more than one-third of the entire training time (left pillar) for GPU on Kaggle and dominates (over 97\%) the entire training time for training on single-node CPU on Terabyte (right pillar). The problem of large tensor traversal magnifies specifically on CPUs. \footnotemark[2]
\footnotetext[2]{We also run single-node GPU training on Terabyte and find scale operation barely surpasses 50\%.} 

We found that periodic updates of the quantization scale effectively eliminate the latency of founding scales. These updates amortize the step's overhead over hundreds or thousands of iterations without hurting model convergence. 
We provide empirical results in Section ~\ref{experiment_res} and in-depth ablation studies in Appendix \ref{period_up} to show that periodic updates can even boost model accuracy. 

\begin{table}[!ht] 
\begin{minipage}{0.4\linewidth} 
\small 
\setlength\tabcolsep{1.pt} 
\captionsetup{justification=centering} 
\caption{DLRM model architecture configurations} 
\label{tab:model_architecture} 
\vspace{-1em} 
\begin{tabular}{@{}ccc@{}} 
\toprule 
Model & \multirow{2}{*}{Kaggle} & \multirow{2}{*}{Terabyte} \\ 
Specifications & & \\ 
\midrule 
\# Embed Tables         & 26                & 26 \\ 
Max \# Row in Tables & 10131227          & 9994101 \\ 
Embed Feature Size     & 16                & 64 \\ 
Bot MLP Arch            & 13-512-256-64-16  & 13-512-256-64 \\ 
Top MLP Arch               & 512-256-1         & 512-512-256-1 \\ 
\bottomrule 
\end{tabular} 
\end{minipage} 
\hspace{5em} 
\begin{minipage}{0.4\linewidth} 
\small 
\captionsetup{justification=centering} 
\caption{DLRM Embedding Tables Quantization, accuracies evaluated on the Kaggle Dataset \label{tab:emb_table}} 
\vspace{-1em} 
\setlength\tabcolsep{1.pt} 
\begin{tabular}{@{}ccc@{}} \toprule 
Weight & \multicolumn{2}{c}{Testing} \\ 
bitwidth & Accuracy & ROC AUC \\ 
\midrule 
FP32 & 78.923\%             & 0.8047 \\ 
\midrule 
INT16     & 78.928\% (+0.005\%)  & 0.8046 (-0.0001) \\ 
INT8      & 78.985\% (+0.062\%)  & 0.8054 (+0.0007) \\

\hc INT4  & \textbf{79.092\% (+0.169\%)} & \textbf{0.8073 (+0.0026)} \\ 
\bottomrule 
\end{tabular} 
\end{minipage} 
\end{table}

\subsection{Distributed Training with Gradient Quantization with only MLP Error Compensation} 
\label{met_distributed} 
We package the entire modified quantized model for DLRM into the Deep Quantized Recommendation Model (DQRM). 
To make DQRM more efficient and competitive in training time compared with normal training, we further optimize the communication workload that occurs in every iteration.

We break down multi-node distributed data parallelism training of DQRM on both GPUs and CPUs in Appendix \ref{additiondistributedtraining}. We found that although PyTorch's built-in specified sparsity greatly reduces all-reduce communication message load, all-reduce gradient communication after every iteration still dominates the training time. The observation motivates us to compress communication further. 
We explore quantizing all gradients into fixed-point INT8 format on top of the applied significantly specified sparsity. However, naively doing so is challenging. Ablation studies presented in Appendix \ref{grad_bit} show that naively quantizing the gradient into INT8 hurts model convergence significantly. Although ~\cite{yang2020training}'s error compensation scheme can prevent accuracy degradation, naively performing error compensation to all parts of the weights can also be highly inefficient due to the enormous error buffer needed for the large embedding tables, as adding large error buffers will greatly increase the model size stored in memory during training. \\ 
\indent We identify surprisingly that gradients in MLP layers are more sensitive to quantization than those in embedding tables while embedding tables are much more robust when specified gradients are heavily quantized, as shown empirically in Appendix \ref{grad_bit}. Therefore, we choose to only compensate the MLP layers for gradient quantization. We achieve reasonable accuracy degradation while compressing the communication message size by roughly 4$\times$ on top of the already heavy and well-supported gradient sparsification added to embedding tables. Detailed experimental settings, results, and evaluations are presented in Section ~\ref{grad_compr}.

%% file: s3_experiments.tex
\section{Experimental Results} ~\label{experiment_res}

\begin{figure*}[t] 
\includegraphics[width=\textwidth]{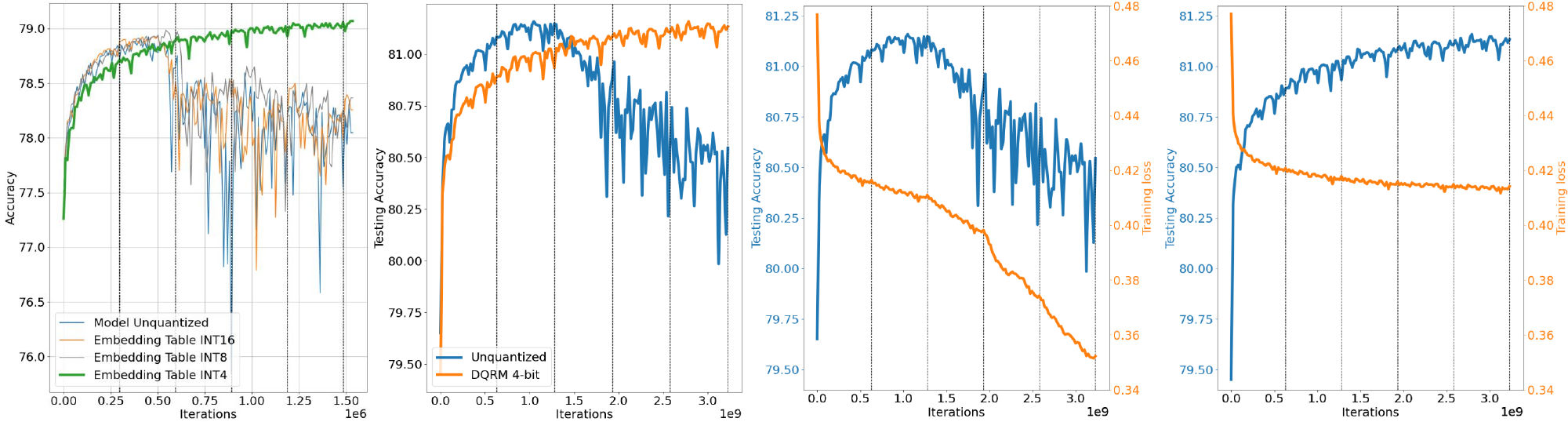} 
\caption{(a) shows the effect of using different QAT bit widths on quantizing embedding tables in DLRM for five epochs of training (epochs are separated by the black dashed lines in all figures). QAT in uniform 4-bit overcomes the severe overfitting suffered by the original DLRM training and leads to significantly higher testing accuracy over five epochs of training. (b) shows the comparison between DQRM 4-bit compared to normal training on the Terabyte dataset; DQRM, with a significantly smaller model size, achieves on-par test accuracy as DLRM FP32 model by better overcoming the overfitting problem. (c) shows that the training loss (orange curve) for normal training starts decreasing drastically in the third epoch, right where the overfitting occurs. In (d), the training loss curve for 4-bit DQRM decreases stably throughout five epochs of training.} 
\label{fig:res_figu} 
\end{figure*}

In this section, we present results evaluating DQRM on two popular datasets for CTR: the Criteo Kaggle Display Advertising Challenge Dataset (shortened below as the Kaggle dataset) and the Criteo Terabyte Dataset (shortened as the Terabyte dataset). 
We followed ~\cite{DLRM19}'s mlperf DLRM optimal configurations for each dataset as summarized in Table ~\ref{tab:emb_table}.

\subsection{Quantization of Embedding Tables} \label{quant_emb_tables}

Embedding tables occupy 99\% of DLRM, which motivates heavy model compression of embedding tables. 
We used one Nvidia A5000 GPU to study the effect of quantization on the embedding tables. Different bit widths for embedding table quantization are used: INT16, INT8, INT4. Unlike previous works ~\cite{guan2019post, deng2021low} that utilize a row-wise scheme for embedding table quantization, we quantize the embedding table with a per-table scheme, as it reduces the number of FP32 scales stored and retrieved every iteration. 

Figure ~\ref{fig:res_figu}(a) illustrates testing accuracy curves for various quantization bit widths. DLRM single precision (blue) overfits after the second epoch, causing a steep accuracy drop, which is a common issue in large-scale CTR models as recently explored in ~\cite{zhang2022towards}. INT16 (orange) and INT8 (grey) follow similar overfitting patterns. However, INT4 (green) converges slower but overcomes overfitting, increasing accuracy over five epochs. In supplementary materials, we empirically show that INT4 convergence lasts up to 16 epochs. We also found that INT2 prevents overfitting like INT4, although it incurs a greater accuracy degradation.

We compare the performance after embedding table quantization in Table ~\ref{tab:emb_table}. 
Uniform INT4 quantization outperforms the original model in testing accuracy by roughly 0.15\% and ROC AUC score by 0.0047 and achieving 8$\times$ reduction in embedding table size. The accuracy improvement is significant in the Kaggle dataset. 
We studied the weight distribution of the training of FP32 weights, INT8, and INT4. We found that the range of distribution for the single-precision weights constantly shifted outwards, while the INT8 weights closely followed this trend. In contrast, INT4 only loosely captures the trend, changing the entire weight distribution much slower compared to higher precision weights. We hypothesize that such observation is linked to DQRM INT4 strong regularization ability. Details are in Appendix \ref{dist_shift}. 

\subsection{Quantization of the Whole Model} \label{quant_whole} 
DLRM models have two components: Embedding tables and MLP layers. 
Compared with the embedding table, we observe that MLP layers are more sensitive to quantization, aligning with ~\cite{deng2021low, zhao2022analysis}. Channel-wise quantization of MLP layers performs better, as it has hardly any accuracy drop from the single-precision MLP, while INT4 matrix-wise MLP quantization badly converges. Additional ablation studies are presented in Appendix \ref{quant_diff_model}. 

We evaluate DQRM INT4 on both the Kaggle and Terabyte datasets. 
We used the same experiment platforms for the Kaggle model as in Section ~\ref{quant_emb_tables}. However, to meet the demanding resource required by the Terabyte models, we use the Intel Academic Compute Environment (ACE) CPU clusters and specifically Intel(R) Xeon(R) Platinum 8280 CPUs for model training and inference. The quantized models are all trained in DQRM for five epochs till convergence. 

\begin{table}[h] 
\caption{DQRM 4-bit quantization results evaluated on Kaggle and Criteo datasets}  
\label{res_tab} 
\vspace{-1em} 
\subfloat[\small 4-bit quantization for DLRM on Kaggle]{ 
\small 
\begin{tabular}{ccccccc} \toprule 
Quantization & Model & \multirow{2}{*}{Model Size} & \multirow{2}{*}{Training Loss} & Training & \multicolumn{2}{c}{Testing} \\ 
Settings & Bit Width &  & & time/it & Accuracy & ROC AUC \\ 
\midrule 
\hc Baseline & FP32 & 2.161 GB & 0.304 & 7 ms & 78.923\% & 0.8047 \\ 
Vanilla PTQ & INT4  & 0.270 GB & -        & - & 76.571\%  & 0.7675  \\ 
\midrule 
PACT*\footnotetext{*PACT~\cite{choi2018pact} uses DoReFa~\cite{zhou2016dorefa} for weight quantization} ~\cite{choi2018pact} & INT4 & 0.270 GB & \multicolumn{4}{c}{Cannot Converge} \\ 
\multicolumn{2}{c}{(MLP in FP32)} & 0.271 GB & 0.303 & 69 ms & 78.858\% & 0.8040 \\ 
\midrule 
LSQ~\cite{esser2019learned} & INT4 & 0.270 GB & 0.350 & 25 ms & 78.972\% & 0.8051 \\ 
\multicolumn{2}{c}{(MLP in FP32)} & 0.271 GB & 0.352 & 21 ms & 78.987\% & 0.8059 \\ 
\midrule 
HAWQ~\cite{dong2019hawq} & INT4 & 0.270 GB & 0.437 & 31 ms & 79.040\% & 0.8064 \\ 
\multicolumn{2}{c}{(MLP in FP32)} & 0.271 GB & 0.436 & 27 ms & 79.070\% & 0.8075 \\ 
\midrule 
\hc DQRM (Ours) & INT4     & 0.270 GB & 0.437 & 22 ms & 79.071\% & 0.8073 \\ 
\hc \multicolumn{2}{c}{(MLP in FP32)} & 0.271 GB & 0.436 & 20 ms & 79.092\% & 0.8073 \\ 
\bottomrule 
\end{tabular}
} 
\centering 
\setlength\tabcolsep{1.pt} 
\subfloat[\small 4-bit quantization for DLRM on Terabyte]{
\small
\begin{tabular}{cccccccccc} \toprule 
Quantization & Model & Model & \multirow{2}{*}{Training Loss} & \multirow{2}{*}{Training time/it} & \multicolumn{2}{c}{Testing} \\
Settings & Bit Width & Size &  &  & Accuracy & ROC AUC \\ 
\midrule 
\hc Baseline & FP32     & 12.575 GB & 0.347071 & 19 ms & 81.165\% & 0.8004 \\ 
\midrule 
Vanilla PTQ & INT4     & 1.572 GB & -        & - & 78.681\% & 0.7283 \\ 
\midrule 
PACT*~\cite{esser2019learned} & INT4 & 1.572 GB & \multicolumn{4}{c}{Cannot Finish $>$1000 ms/it} \\ 
\midrule 
HAWQ~\cite{dong2019hawq} & INT4 & 1.572 GB & \multicolumn{4}{c}{Cannot Finish $>$1000 ms/it} \\ 
\midrule 
LSQ~\cite{esser2019learned} & INT4 & 1.572 GB & 0.350 & 42 ms & 81.134\% & 0.7996 \\ 
\multicolumn{2}{c}{(MLP in FP32)} & 1.572 GB & 0.356 & 42 ms & 81.127\% & 0.7998 \\ 
\hc DQRM (Ours) & INT4        & 1.572 GB & 0.409774 & 29 ms & 81.210\% & 0.8015 \\ 
\hc \multicolumn{2}{c}{(MLP in FP32)} & 1.572 GB & 0.412 & 29 ms & 81.200\% & 0.8010 \\ 
\bottomrule 
\end{tabular}}
\end{table} 

Figure ~\ref{fig:res_figu} (b), (c), and (d) display training loss and testing accuracy curves for the Terabyte dataset. In (b), the original model testing accuracy (blue) overfits at the 2nd epoch, while DQRM INT4 (orange) steadily rises and avoids overfitting. Comparing DLRM (c) and DQRM INT4 (d), the latter demonstrates better regularization, with a consistent decrease in loss (orange curve in (d)) and a plateau in testing accuracy (blue curve in (d)). In contrast, DLRM's accuracy (blue curve in (c)) and training loss (orange curve in (c)) crash after the second epoch.

We report single-node training and testing performance in Table ~\ref{res_tab}. We compare DQRM INT4 against PTQ and prior popular QAT works ~\cite{esser2019learned, choi2018pact, dong2019hawq} that achieve strong INT4 results on CNN. Unfortunately, ~\cite{deng2021low} does not open-source their INT4 PTQ implementation, so we apply vanilla PTQ using our quantization method with channel-wise quantization for MLP. Also, we implemented ~\cite{esser2019learned, choi2018pact} ourselves on DLRM. 
Table ~\ref{res_tab} (a) shows the experimental results on the Kaggle dataset using GPU. 
Using an update period of 200, DQRM achieves both a higher testing accuracy (79.071\%) and a higher testing ROC AUC score (0.8073) than DLRM in FP32. 
On the Terabyte dataset in (b), by using an update period of 1000, DQRM INT4 achieves slightly higher than FP32 baseline test accuracy (0.45\%) and testing ROC AUC (0.0011). The vanilla PTQ incurs a significant accuracy drop in both datasets. Interestingly, prior QAT works exhibit their inefficiencies. PACT~\cite{choi2018pact} and LSQ~\cite{esser2019learned} do not avoid overfitting completely like DQRM and, thus, achieve lower testing accuracy on both datasets. HAWQ~\cite{dong2019hawq} cannot effectively eliminate unnecessary memory traversal, making it consistently slower than DQRM in training time and even failing to finish on Terabyte. 

\begin{table*}[!ht] 
\caption{Communications compression for Distributed Data Parallelism training among four nodes or GPUs, the baseline used here has forward prop weights in 4-Bit} 
\vspace{-1em} 
\label{tab:compress_communication} 
\centering 
\begin{minipage}{1.0\textwidth} 
\raggedright 
\centering 
\small 
\setlength\tabcolsep{1.pt} 
\begin{tabular}{@{}cclccccc@{}} \toprule 
Model & Training & Communication        & Communication     & Latency  & Training & \multicolumn{2}{c}{Testing} \\ 
Settings & Platforms & Compression settings & Overhead per iter & per iter & Loss     & ACC & ROC AUC          \\ 
\midrule 
\multirow{4}{*}{Kaggle} & 4X & grad uncompressed & 2.161 GB & $>$1000 ms & 0.436 & 78.897\%\footnotemark[1] 
& 0.8035 
\\
& Nvidia A5000 & \hspace{3mm} + EMB gradient sparse\footnotemark[2] 
                      & 2.010 MB & 61 ms & 0.436 & 78.897\% & 0.8035\\ 
& GPUs & \hspace{3mm} + INT8 grad Quantization 
& 0.509 MB & 110 ms\footnotemark[3] & 0.442 & 78.840\% & 0.8023 \\ 
\midrule 
\multirow{4}{*}{Terabyte} & 2X (2 processes) 
& grad uncompressed & 12.575 GB   & $>$1000 ms     & 0.412 & 81.156\% & 0.7997 \\
& Intel(R) Xeon(R) & \hspace{3mm} + EMB grad sparse           & 6.756 MB   & 210 ms     & 0.412 & 81.156\% & 0.7997 \\ 
& Platinum 8280 CPU & \hspace{3mm} + INT8 grad Quantization & 1.732 MB    & 225 ms      & 0.414 & 81.035\% & 0.7960 \\ 
\bottomrule 
\end{tabular} 
\begin{flushleft} 
\footnotetext[1]{Data parallelism consistently lowers the test accuracy in all settings compared with single-node training.} 
\footnotetext[2]{Sparsification used is specified sparsity which is a lossless compression for embedding tables so the testing accuracy is exactly the same as uncompressed case.} 
\footnotetext[3]{PyTorch sparse tensor \textit{allreduce} library doesn't support low-precision arithmetic, without further system-level effort in low-precision optimization, the latency per iteration increases purely from the quantization overhead per iteration.} 
\end{flushleft} 
\end{minipage} 
\end{table*} 

\subsection{Adding Gradient Quantization on Specified Sparsity} 
\label{grad_compr}
Our experiments are based on synchronous data parallelism (DP). However, we argue that the gradient compression techniques presented are directly applicable to hybrid parallelism settings as in ~\cite{DLRM19}. Moreover, the only difference between the data parallelism and the DLRM's hybrid parallelism is the all2all round of communication, and it can also be directly benefited from DQRM INT with message load shrunk by 8$\times$ directly. Due to PyTorch \texttt{DistributedDataParallel} library limitations, we customized and open-sourced our own Data Parallel library implementing gradient quantization before \texttt{allreduce}. Appendix \ref{real_dist} details the implementation. Our approach supports multi-GPU and CPU node platforms, with gradient compression experimental results in Table ~\ref{tab:compress_communication}. 

For experiments on the Kaggle dataset, we utilized four NVIDIA A5000 GPUs. For the Terabyte dataset, we ran on two Intel(R) Xeon(R) Platinum 8280 CPU nodes. Naively, because of the giant model size, when the gradient is completely uncompressed, the overhead of gradient communication is up to 2.1 GB per iteration. 
PyTorch has built-in support to exploit specified sparsity in the EmbeddingBag modules which is very powerful in the \textit{allreduce} in DP and can compress 2.1 GB to 2 MB. 
Moreover, using specified sparsity is lossless and has no impact on model training performance.

Surprisingly, after the specified sparsity is applied, the MLP gradients become significant among the total gradient left. After careful profiling, they occupy around 95\% and 45\% in size among the remaining gradients for training in DP on Kaggle and Terabyte respectively. 
We further add error compensation to the MLP layers to reduce accuracy degradation. 
PyTorch doesn't support low-precision sparse \textit{allreduce} well. With limited backend support and significant communication overhead caused by sparse tensor coalescing, achieving speedup from communication reduction remains difficult. 
With in-depth system-level optimization for low-precision sparse \textit{allreduce} and gradient quantization workload, the training time cost reduction of distributed DQRM can be realistically fulfilled, but such an endeavor is outside the scope of our current work. Nevertheless, we showed empirically that adding gradient quantization only introduces a trivial decrease of 0.057\% in the testing accuracy and 0.0012 in the ROC AUC score despite a significant reduction in communication size for the Kaggle dataset. 

Similarly, gradient compression of the embedding tables introduces insignificant accuracy loss with roughly 0.1\% for testing accuracy and less than 0.004 for testing ROC AUC. 
We detail more evaluations of MLP gradient sensitivity for quantization and the effect of error compensation in Appendix \ref{grad_bit}. 

%% file: s6_conclusion.tex
~\section{Conclusion} \label{conclusion} 
In this work, we propose a systematic quantization framework DQRM for large-scale recommendation models. Specifically, we discover that the DLRM model suffers severely from the overfitting problem. We show that ultra-low precision quantization can help overcome the strong overfitting and better utilize the training dataset, which eventually leads to higher test accuracy. We observe that conventional QAT is troublesome in training large-scale recommendation models and we propose two techniques that significantly alleviate the issue. Besides, to further optimize DQRM under the distributed environment, we combine specified sparsification and quantization together to compress communications. Our framework is intensively evaluated on the published dataset Kaggle and Terabyte, where we outperform the full-precision DLRM baselines while achieving an 8$\times$ reduction of model size.

%% file: s7_supplements.tex
\section*{Appendix} 
\maketitle
\input{s4_ablation.tex}

\section{Experiment Platforms} 
\subsection{Distributed Environment with Gradient Quantization} 
\label{real_dist} 

\begin{figure}[h] 
\centering 
\includegraphics[width=0.5\textwidth]{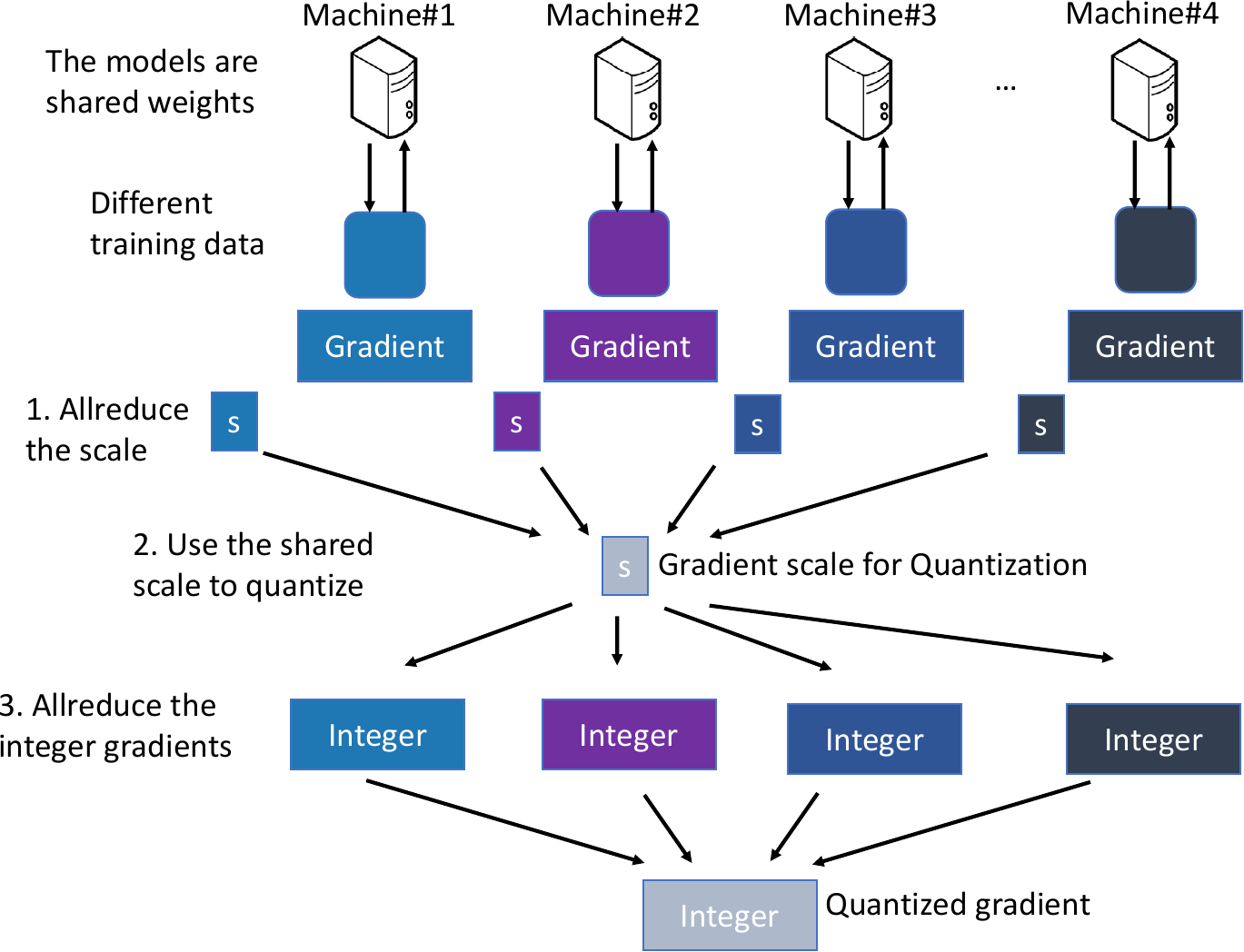} 
\caption{Illustration of the Customized Data Parallelism Framework that Supports Gradient Quantization} 
\label{illusq} 
\end{figure} 

In the PyTorch DistributedDataParallelism library, during backward propagation, the error tensor's backward() function handles both backpropagating gradients to local parameters together with the \textit{allreduce} of gradient tensor across nodes. Therefore, after the powerful single line of code, gradients behind each parameter tensor in the distributed data parallelism model have their corresponding gradient ready. However, the quantization of the gradient has to happen between the local backpropagation and the global \textit{allreduce}. Therefore, the distributed data parallelism package is not very handy. 

To implement gradient quantization, we implement a stand-alone framework that performs the majority of distributed data parallelism workflow. To implement gradient quantization, we summarized our implementation in Figure ~\ref{illusq}. We start by partitioning the training batch into pieces to be used by different machines. After forward propagation, the local devices are asked to backpropagate the corresponding gradient locally. Then, the following sequence of steps is performed for all parameters in the model. The quantization scale of parameter tensors is computed locally at first. The quantization scale is completely determined by the quantization bit-width and local tensor distribution so can be performed locally. Then, the quantization scale is first allreduced to be unified. Then, using the same scale, gradient tensors on each device are quantized locally. The integer gradients are then allreduced. After two \textit{allreduce} steps, each device then has the same quantization scale and the same quantized gradients. The local step update of parameters then can be dequantized. The quantization bit width used is INT8 and unified across all parameters. When it comes to sparse gradients from embedding tables, we only quantize the \texttt{values} part of the \texttt{sparse\_coo} tensor format and leave the \texttt{indices} and other fields unchanged. 

One caveat of the entire flow of our customized gradient communication is that it can only be used for \textit{allreduce} paradigms that perform the reduced operation at the destined machine, such as the Ring allreduce. However, for recursive doubling \textit{allreduce}, where the reduced operation is collectively performed by many machines, our framework is not applicable. The main problem is at each reduced operation, where a sum is taken by two low-precision operations. To continue propagating the gradients under the low-precision bit-width, an additional rounding should occur, which may potentially lose the gradient precision further. Unfortunately, PyTorch \textit{allreduce} operations use recursive-doubling \textit{allreduce}, throughout our presented experiment, we use FP32 tensors to contain quantized INT8 integer gradients to avoid accumulation and rounding issues from the impact of \textit{allreduce}. 

\subsection{Additional Analysis on DLRM Distributed Training} 

\label{additiondistributedtraining} 

Firstly, we provide more details on the experiment setup for Figure 1(c). The experiments are run on both a single GPU and a single CPU node. We train the model under the Kaggle dataset settings on a single Nvidia A5000 GPU, as shown in the left pillar. We also run experiments of the Terabyte dataset on a single Intel(R) Xeon(R) Platinum 8280 CPU node. We found that such a problem significantly magnifies on the CPUs. Finding the quantization scale for large DLRM models under the Terabyte dataset settings can take more than 1000 ms per iteration of QAT while occupying the majority of the training time. The Terabyte dataset has embedding tables that are ten times larger than those on the Kaggle dataset. \\ 

\begin{figure}[h!] 
\centering 
\includegraphics[width=0.7\linewidth]{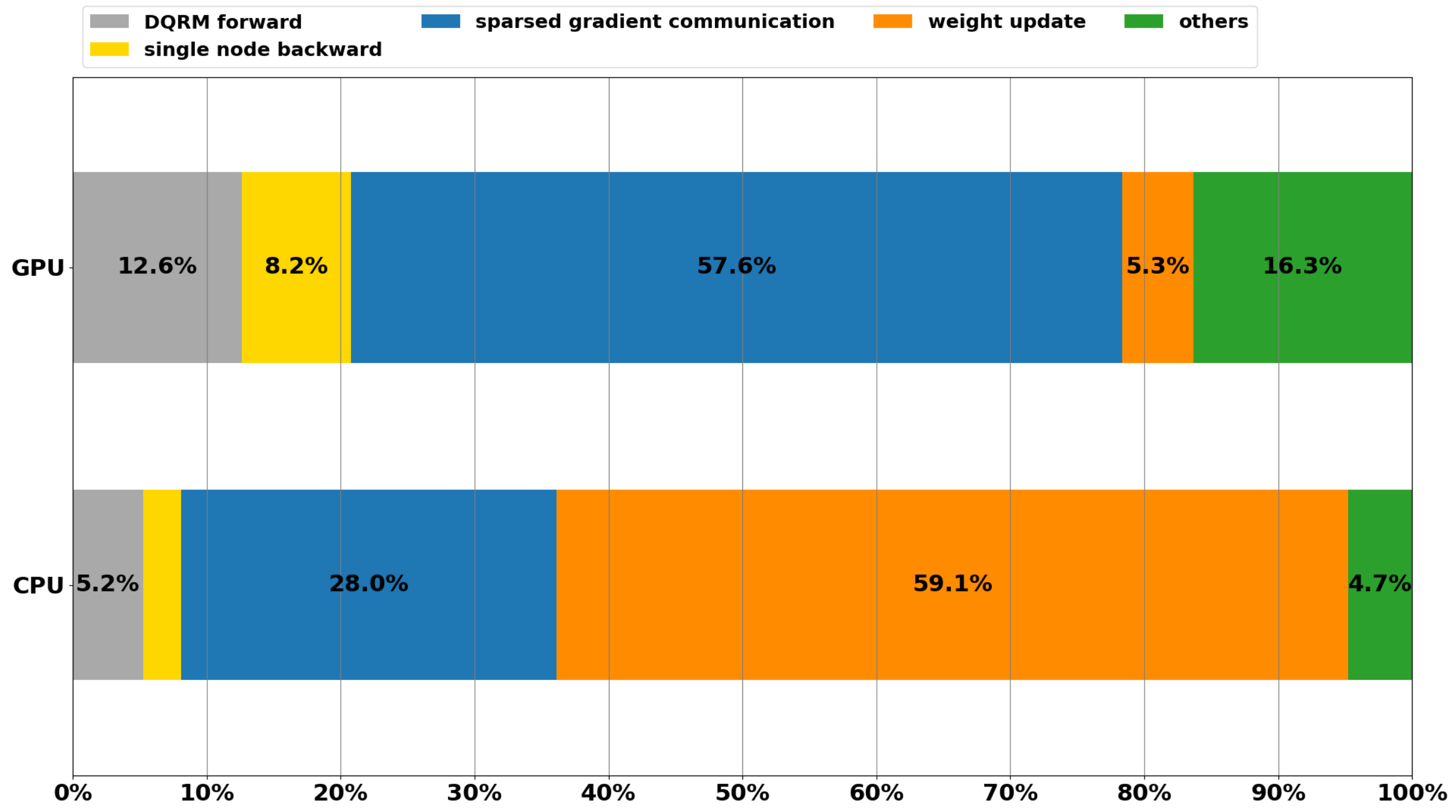} 
\caption{A breakdown of training time on distributed data parallelism environment on multiple GPUs and CPU nodes} 
\label{fig:training_time_breakdow} 
\end{figure} 

In Figure ~\ref{fig:training_time_breakdow}, we profile the total time spent during the entire training process when using the PyTorch built-in specified sparsity on Embedding tables. During the profile, we train DQRM Kaggle across four Nvidia A5000 GPUs and DQRM Terabyte across two Intel Xeon Platinum 8280 CPUs with an update period set to 1000 to eliminate the difficulties of finding the scale. Under the pure Data Parallelism settings, gradient communication (blue) is a severe bottleneck for both GPU and CPU platforms. Notice that the weight update (orange) portion of CPU platforms is significant, because the vanilla PyTorch library we used is highly inefficient in sparse tensor operations, and the inefficiency in arithmetic will be largely reduced by the recently published IPEX\footnote{IPEX official GitHub repo: https://github.com/intel/intel-extension-for-pytorch} package from Intel. Besides the significant weight update (orange) portion which can be optimized relatively easily, we argue that the gradient communication (blue) is still a severe bottleneck on the CPU platform. 

\subsection{Simulated Data Parallelism Environment} 

\begin{algorithm} 
\caption{Simulated Data Parallelism on single machine} 
$\texttt{buffer\_clean} \gets False$\; 
\For{\texttt{j}, \texttt{batch} in enumerate(\texttt{train\_loader})} {
    $\texttt{Z} \gets $ model\_forward(\texttt{training\_input})\; 
    $\texttt{E} \gets $ loss\_criterion(\texttt{Z}, \texttt{label})\; 
    clear\_gradient(\texttt{model})\; 
    \If {\texttt{buffer\_clean}}{
        grad\_buffer\_zero()\; 
        $\texttt{buffer\_clean} \gets $ False\; 
    }
    \texttt{E}.backward() \Comment*[r]{local backward propagation} 
    grad\_buffer\_update()\; 
    \If {\texttt{j} \% \texttt{simulated\_nodes} == 0}{
        weight\_buffer\_update()\; 
        $\texttt{buffer\_clean} \gets $ True\; 
    }
} 
\label{algo1} 
\end{algorithm} 

For some experiments presented in the main content, we use a simulated environment to run Data Parallelism on single device to imitate the multiple device environment. Here we provide more details of our implementation. In our implementation, we add an addition gradient buffer for every parameter tensor. As summarized in Algorithm~\ref{algo1}, the batch size is splited similar to one distributed machine. In the second if statement, the weight is actually updated by the gradient every simulated\_node iterations. Between updates, training still backpropagate gradients from relevant iteration and are hold inside the parameter buffers. After weight being updated by the buffer, the buffer is cleared prior to new gradient being backpropagated. One key difference between simulated and the real distributed environment lays in its \textit{allreduce} mechanism. Shown in Section ~\ref{real_dist}, if gradients are quantized, two \textit{allreduce} processes occur in one iteration: the first communicates quantization scale, while the second one communicates quantized gradient tensors. However, two \textit{allreduce} processes are difficult to implement in the simulated environment. Therefore, we reuse the quantization scale of the first iteration throughout simulated\_machine number of iterations in our implementation, which can potentially hurts the gradient quantization precision. However, through our experiments, we didn't observe this case.

\subsection{Simulated Framework Evaluation} 
\begin{table}[h] 
\caption{Multi-node Experiment Results with 8-bit gradients, loss evaluated on the Terabyte Dataset} 
\label{tab:multinode} 
\centering 
\begin{adjustbox}{width=0.45\textwidth}  
\centering 
\small 
\setlength\tabcolsep{1.pt} 
\begin{tabular}{cccc} \toprule 
\multirow{2}{4em}{\#Node} & Training Loss & \multicolumn{2}{c}{Testing} \\
& Drop & Acc Drop & AUC Drop \\ 
\midrule 
2 & \textbf{-0.002447} & 0.092 & \textbf{0.0025} \\ 
4 & -0.00273  & 0.114 & 0.0036 \\ 
8 & -0.00395  & \textbf{0.059} & 0.0053 \\ 
\bottomrule 
\end{tabular}
\end{adjustbox} 
\end{table} 
We also evaluate the effect of different node counts on gradient quantization. The result is listed in Table ~\ref{tab:multinode}. Currently, different node counts are simulated on the single CPU node. Across three different node counts, 2, 4, and 8, the drop in training loss, testing accuracy, and ROC AUC score is consistent and small. 

\section{Demo Recommender on the Phone}
\begin{figure*}[t] 
\includegraphics[trim=0cm 10cm 0cm 0cm, scale=0.53]{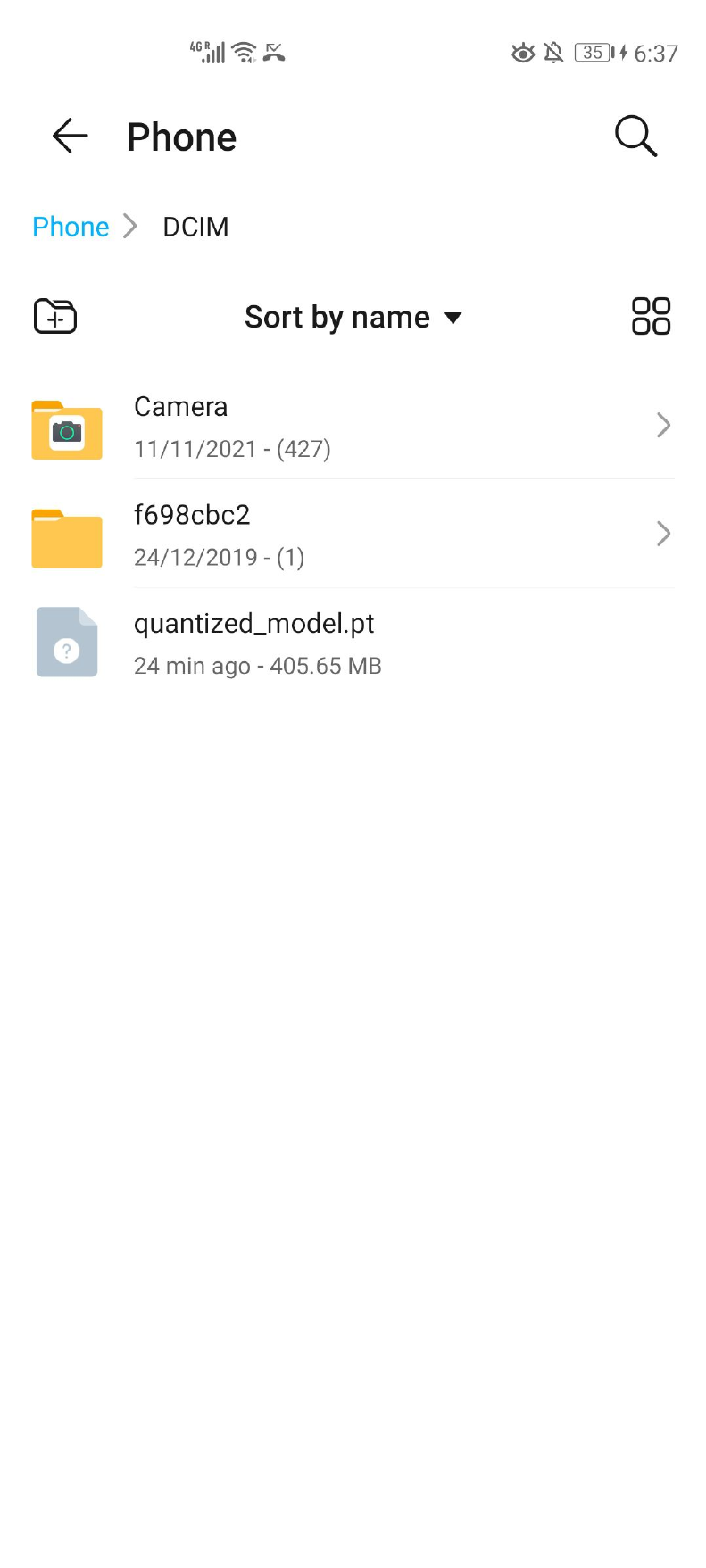} 
\caption{A screenshot of the quantized model on a high-end Android Phone} 
\label{fig:phone}
\end{figure*}

We include a demo of DQRM exported to an Android phone. From Figure~\ref{fig:phone}, the tested DQRM model size is 405.65 MB. As a reference, the DLRM Kaggle model size is 2.16 GB. The model size is not strictly 8$\times$ compression because of the following two reasons: 1) Embedding tables can be quantized into INT4, but the embedding vectors have to be bit-packed together into INT8 format to fully benefit from the INT4 low precision. However, bitpacking on PyTorch is tricky and PyTorch 4-bit packing is not fully optimized in terms of performance. 2) PyTorch doesn't support MLP layers to be quantized into bitwidth below INT8. Therefore, MLP layers, although quantized into INT4, still need to be stored as INT8 numbers. Still, we believe this is a first step towards deploying large recommendation models to edge devices so as to alleviate the heavy cloud AI inference pressure.

%% file: s4_ablation.tex
\section{Ablation Studies} 
\label{abl_st} 
In the ablation studies, we present extended studies on the parameters of the DQRM framework, including different update period's impact on DQRM in Section ~\ref{period_up}, different gradient quantization bit width's effect on model convergence in Section ~\ref{grad_bit}, different sensitivity to quantization among layers in DQRM in Section ~\ref{quant_diff_model}, and the effect of QAT without retraining on DQRM. Then, we evaluate DQRM INT4 with the large number of epochs in training in Section ~\ref{qat_no_pret}. 

\subsection{Periodic update} \label{period_up} 

As shown in the methodology section, the sheer size of embedding tables renders table traversal a bottleneck during the forward pass. We propose to utilize periodic updates of the quantization scale of each embedding table during quantization in order to amortize the huge latency of table lookup into multiple iterations. Among the options, we found that the optimal period is different for training settings with different model sizes and batch sizes. The experimental results are presented in Table ~\ref{tab:periodic_update}. 

\begin{table}[ht] 
\caption{Evaluation of Periodic Update on Kaggle and Terabyte Datasets} 
\label{tab:periodic_update} 
\centering 
\begin{adjustbox}{width=0.5\linewidth} 
\centering 
\small 
\setlength\tabcolsep{1.pt} 
\begin{tabular}{ccccc} \toprule 
\multirow{2}{4em}{Model Settings} & \multirow{2}{4em}{Period} & Latency & \multicolumn{2}{c}{Testing} \\ 
& & per iter & Accuracy & ROC AUC \\ 
\midrule 
\multirow{3}{4em}{Kaggle}   & 1         & 31 ms & 79.040\% & 0.8064 \\ 
                            & \cellcolor[HTML]{FAD5A5} 200 & \cellcolor[HTML]{FAD5A5} 22 ms & \cellcolor[HTML]{FAD5A5} 79.071\% & \cellcolor[HTML]{FAD5A5} 0.8073 \\ 
                            & 500       & 22 ms & 79.034\% & 0.8067 \\ 
\midrule 
\multirow{3}{4em}{Terabyte} & 1         & $>$1200 ms & - & - \\ 
                            & 200       & 58 ms & 81.159\% & 0.7998 \\ 
                            & 500   & 51 ms & 81.193\% & 0.8009 \\ & \cellcolor[HTML]{FAD5A5} 1000 & \cellcolor[HTML]{FAD5A5} 46 ms & \cellcolor[HTML]{FAD5A5} 81.210\% & \cellcolor[HTML]{FAD5A5} 0.8015 \\ 
\bottomrule 
\end{tabular}
\end{adjustbox} 
\end{table} 

For experiments running on the Kaggle dataset, we run on one Nvidia A5000 GPU. We again trained the model for 5 epochs. During DQRM, all weights are quantized into INT4. 
Training time per iteration decreases significantly, from over 31 ms to 22 ms when the quantization scale is updated once every 200 iterations compared with every iteration. Also, along with decreased single-node training time, model convergence even slightly improved. We also test the update period of 500, and we found diminishing gains in training time reduction and a slight decrease in testing performance. 

Similar experiments are conducted on a single Intel Xeon platinum 8280 CPU on Terabyte. 
We found that with the current ML framework, CPUs have much greater difficulties in traversing large embedding tables. A single iteration of training of the model under the Terabyte setting takes more than 1.2 seconds to finish, thus making it impractical to complete. When the quantization scale is periodically updated once every 200 iterations, the training time per iteration dramatically improves to 57 ms per iteration. Due to the difference in model size and training mini-batch size, we found that using a longer period of 500 iterations brings further speedup in training time per iteration, dropping to 34 ms, while even slightly benefiting the training accuracy (0.03\% compared with 200 iterations). We hypothesize that the reason for the slight accuracy boost is that, QAT inherently has an unstably growing quantization scale, and periodic update of the quantization scale helps stabilize training, which further helps the model converge. 

\begin{figure}[t] 
\centering 
\includegraphics[width=0.8\textwidth]{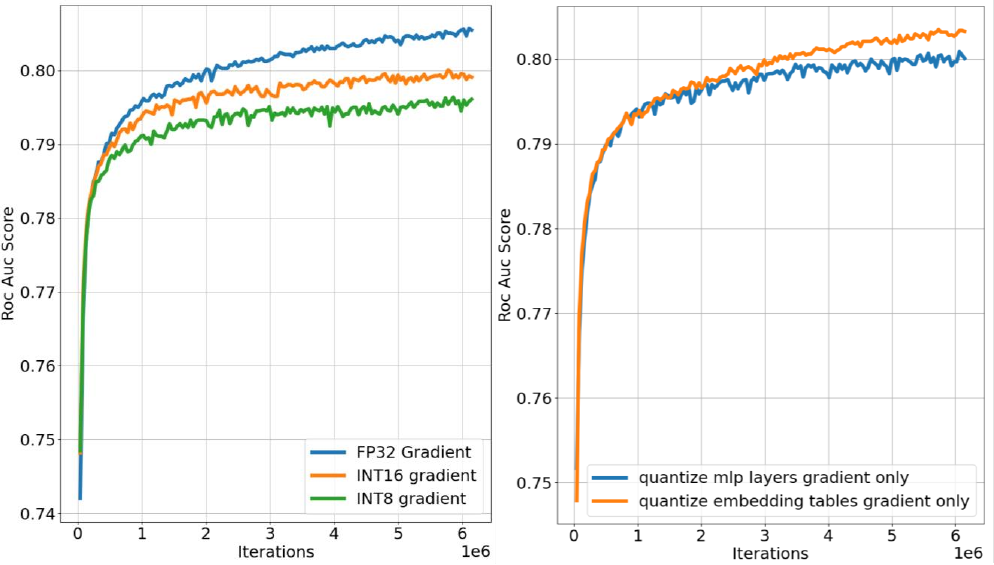} 
\caption{(a) Testing accuracy of naively quantizing gradients into different bit widths. Naive gradient quantization leads to a significant accuracy drop. (b) MLP gradients are more sensitive to quantization. If only quantizing the embedding table gradients into INT16, it will lead to less drop in accuracy compared to only quantizing MLP gradients into INT16.} 
\label{fig:overf_and_sep_quanti} 
\end{figure} 

\subsection{Different Gradient Quantization Bit Width} \label{grad_bit}

In this Section, we present our findings related to the quantization of the gradients during distributed DP training for DQRM. All experiments presented in this Section assume that specified sparsity is enabled. Firstly, we study the effect of naive quantization of gradients on the model convergence. The experiments are run on a single NVIDIA M40 GPU under a simulated DP environment under the Kaggle dataset. More details of the experiment platform setup can be found in \ref{real_dist}. Quantizing all the gradients naively introduces huge accuracy loss and largely hurts the model convergence. In Figure ~\ref{fig:overf_and_sep_quanti} (a), the blue curve signifies the baseline: FP32 gradients. The orange curve is for INT16 gradients, while the green curve is for INT8 quantization gradients. 
Both the INT16 and the INT8 quantization force the training to plateau earlier and converge to a much lower testing accuracy. 

Following our first experiment and setup, we investigate quantizing just embedding table gradients or just MLP layers gradient but not both Figure \ref{fig:overf_and_sep_quanti} (b). We show that only quantizing embedding table gradients allows the model to continuously converge throughout the five epochs while quantizing the MLP layers significantly slows down the model convergence. Therefore, we decided to add Error Compensation only to MLP layers' gradients, as MLP layers' gradients are more sensitive to quantization. More experiments are presented in Table ~\ref{mlp_ec} on the similar 4 Nvidia A5000 GPUs with our custom framework. We showed that quantizing Embedding Table gradient into INT8 doesn't hurt test accuracy much, and MLP error compensation largely prevents the large accuracy degradation from MLP gradient quantization.

\begin{table}[!ht]  
\caption{Evaluation of MLP Quantization and Error Compensation} 
\label{mlp_ec} 
\centering 
\begin{adjustbox}{width=0.7\linewidth} 
\centering 
\small 
\setlength\tabcolsep{1.pt} 
\begin{tabular}{lccc} \toprule 
Settings & Testing Accuracy & Testing ROC AUC \\ 
\midrule 
baseline & 78.897\% & 0.8035 \\ 
Only Embedding gradient in INT8 & 78.858\% & 0.8023 \\ 
Embedding and MLP gradient in INT8  & 78.608\% & 0.7974 \\ 
\hc Embedding and MLP gradient in INT8 + MLP EC & 78.840\% & 0.8023 \\ 
\bottomrule 
\end{tabular}
\end{adjustbox} 
\end{table}

\subsection{Quantization of Different Part of Models} \label{quant_diff_model} 

\begin{table}[!ht]  
\caption{Quantization Evaluation of Each Part of the Model} 
\label{mlp_quantization} 
\centering 
\begin{adjustbox}{width=0.6\textwidth} 
\centering 
\small 
\setlength\tabcolsep{1.pt} 
\begin{tabular}{lccc} \toprule 
Settings & Testing Accuracy & Testing ROC AUC \\ 
\midrule 
Baseline                                                           & 78.718\% & 0.8001 \\ 
\midrule 
\hc + Embedding Tables in 4-bit                                    & 78.936\% & 0.8040 \\ 
\hspace{3mm} + MLP in 4-bit matrix-wise 
                                                                   & 78.830\% & 0.8022 \\ 
\hc \hspace{3mm} + \textbf{MLP in 4-bit channelwise}     & 78.897\% & 0.8035 \\ 
\midrule 
\hspace{3mm} + MLP in 8-bit channelwise                            & 78.950\% & 0.8045 \\ 
\bottomrule 
\end{tabular}
\end{adjustbox} 
\end{table} 

In DQRM training, we observe that different layers exhibit different sensitivities to quantization. To study this, we utilize four NVIDIA M40 GPUs and conduct experiments on the Kaggle dataset under the distributed DP settings. The experiment results are presented in Table ~\ref{mlp_quantization}. Compared with quantizing embedding tables, MLP layers quantization cannot benefit from the diminishing overfitting effect. Instead, when quantizing MLP layers in the matrix-wise fashion into INT4 and quantizing activation during QAT, the DQRM framework fails to converge. When just quantizing the MLP layers in the matrix-wise fashion and without quantizing the activations, DQRM gives more than 0.1\% testing accuracy drop, which is a lot for the Kaggle dataset. In contrast, we analyze MLP quantization in a channel-wise fashion, specifically finding a scale for every row inside the linear layer weights. MLP channel-wise quantization outperforms matrix-wise, bringing the overall accuracy loss for the entire INT4 model to be under 0.04\% and 0.0005 in testing ROC AUC compared to a single precision unquantized MLP model. We also provide the performance of quantizing the MLP layers in INT8 channel-wise, it is on par with the single precision MLP in test accuracy and test ROC AUC score. 

\subsection{QAT without Pre-training} 
\label{qat_no_pret} 

\begin{table}[!ht]  
\caption{Evaluation of QAT from scratch on Recommendation Models} 
\label{qat_finetune} 
\centering 
\begin{adjustbox}{width=0.7\textwidth} 
\centering 
\small 
\setlength\tabcolsep{1.pt} 
\begin{tabular}{lccc} \toprule 
Settings & Testing Accuracy & Testing ROC AUC \\ 
\midrule 
One epoch pretraining + Four epochs of QAT & 78.926\% & 0.8039 \\ 
\hc Five epochs of QAT without pretraining & 78.936\% & 0.8040 \\ 
\bottomrule 
\end{tabular}
\end{adjustbox} 
\end{table} 

Previously, QAT has been used as a fine-tuning technique for the quantization of CNN~\cite{dong2020hawq, zhang2023qd} and transformer models~\cite{shang2023pb, li2023qft}. Usually, with pretraining on the single-precision bit-width, the trained weights can reduce accuracy loss when fine-tuning on low-precision settings. We examine such paradigms on the DLRM models. We train DLRM in QAT but only quantize the embedding tables into INT4 under the Kaggle dataset using four NVIDIA M40 GPUs under the distributed DP settings. We compare DLRM with one epoch of pretraining in the single precision followed by four epochs of INT4 QAT with DLRM with INT4 QAT from scratch. 

We plot the testing accuracy versus iteration curves in Figure ~\ref{fig:overf_and_sep_quanti} (a). The experiment results are presented in Table ~\ref{qat_finetune}. In the diagram, the vertical dashed lines signify the boundary of each epoch. The blue curve is for QAT with pretraining, while the orange curve is without. After the first epoch, we can see that as the model transitions from single-precision to INT4 quantized data type, the blue curve drops slightly, which is expected. Further, in the mid-third epoch, QAT with pretraining (blue) reaches its peak and then drops quickly afterward. QAT from scratch eventually has a slight testing accuracy edge over QAT with pretraining. Also, in Figure ~\ref{fig:overf_and_sep_quanti} (b), we plot the training loss across five epochs for two different settings. QAT with pretraining (blue) has training loss constantly below QAT from scratch. From here, we argue that QAT with pretraining on DLRM models does speed up the model convergence, taking less number of epochs to reach its peak. However, it suffers from earlier overfitting compared with QAT from scratch. Under our settings, we observe that QAT from scratch slightly gains in testing accuracy of 0.01\% with 0.0001 of testing ROC AUC score. 

\begin{figure}[h] 
\centering 
\includegraphics[width=0.8\textwidth]{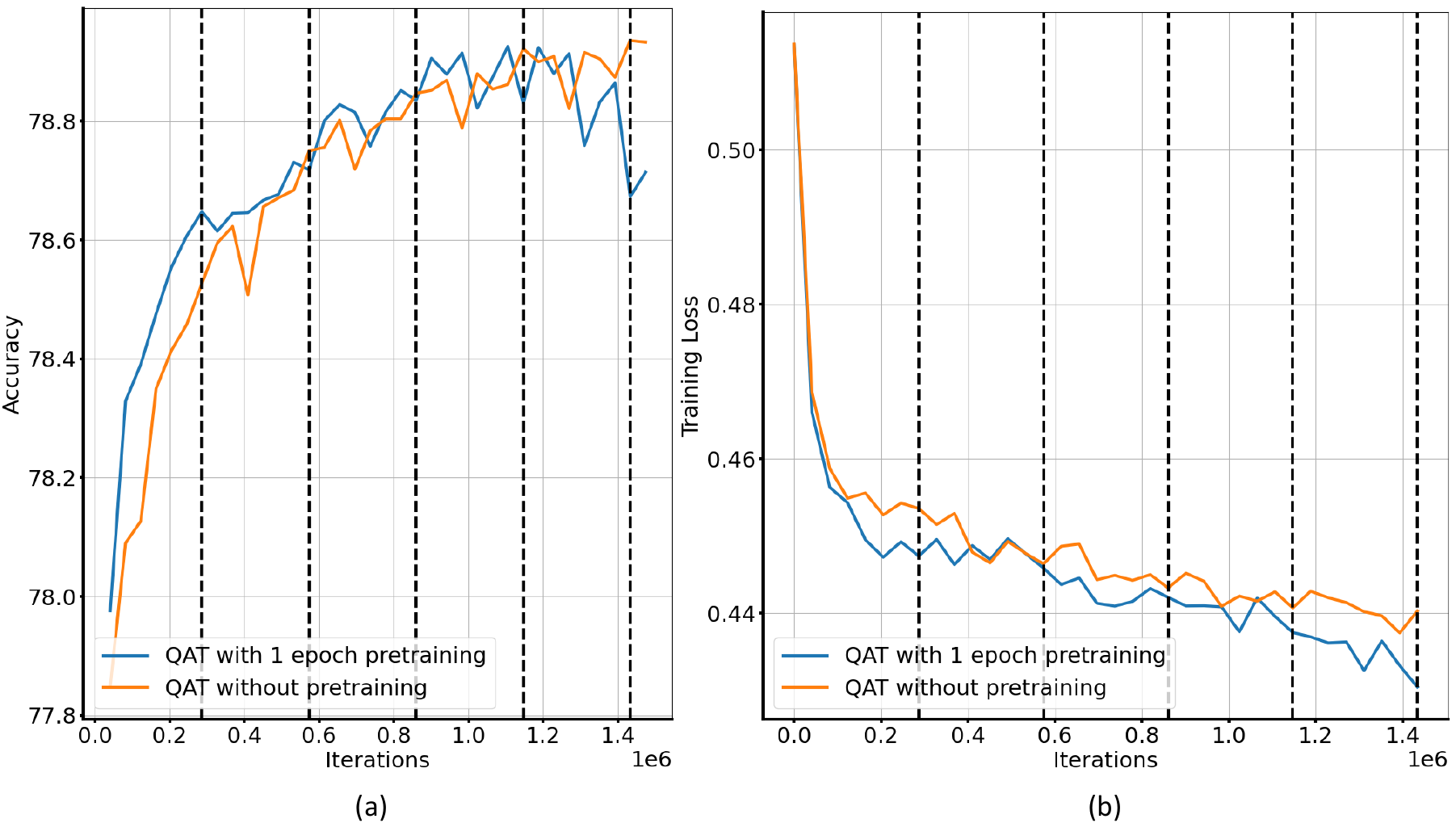} 
\caption{(a) Testing accuracy over the 5 epochs for 1 epoch of pretraining before 4 epochs of QAT and 5 epochs of QAT without pretraining. Pretraining leads to faster overfitting, and QAT without pretraining avoids overfitting and achieves better testing accuracy from 5 epochs of training. (b) Training loss over the 5 epochs. Pretraining before QAT leads to a faster decrease in training loss in DLRM compared with QAT without pretraining.} 
\label{fig:overf_and_sep_quanti} 
\end{figure} 

\subsection{How Quantization Affects Weight Distribution} 
\label{dist_shift} 
\begin{figure*}[t] 
\includegraphics[width=\textwidth]{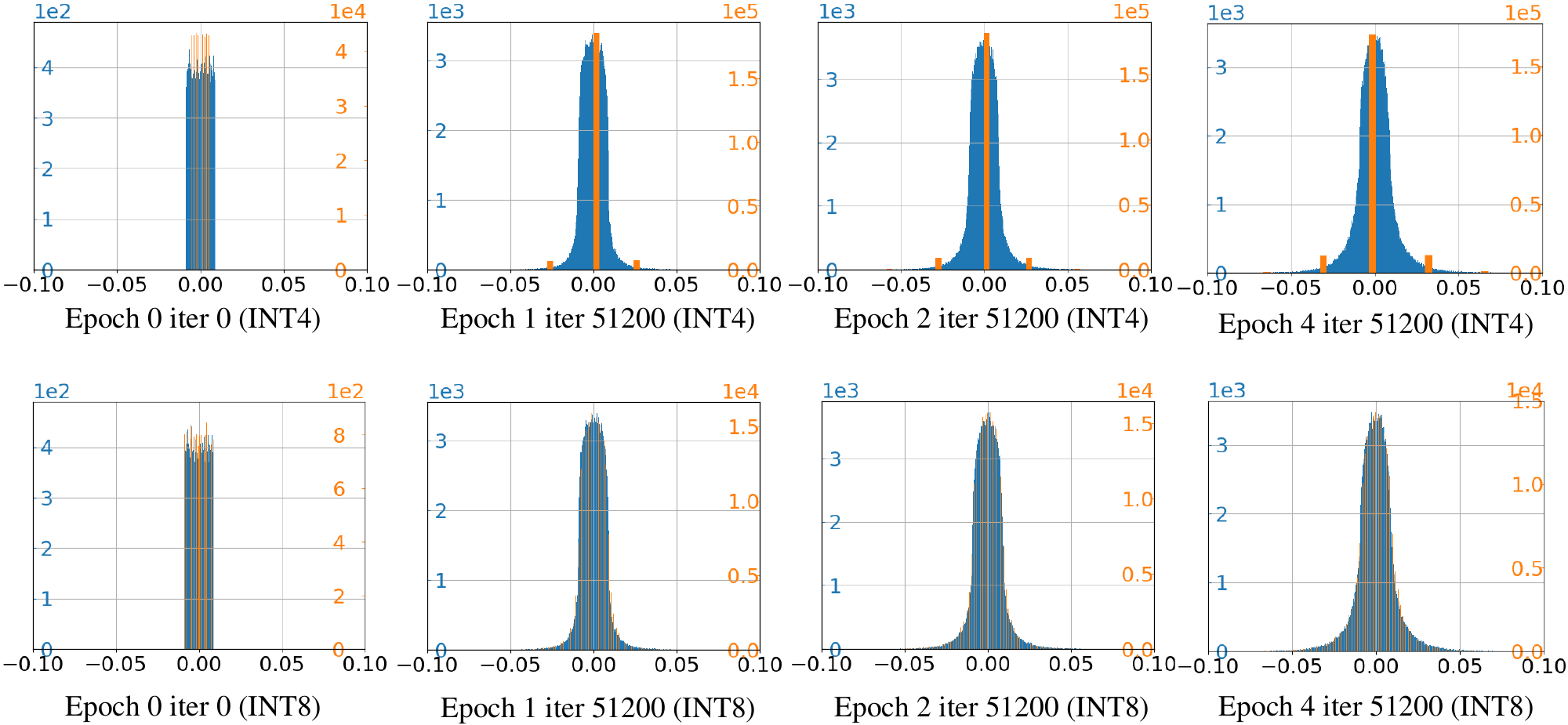} 
\caption{Contrast between INT4 and INT8 quantization-aware training weight distribution shift. Orange pillars are for the quantized weight distribution under QAT with low precision, while blue is for the unquantized weight under normal training. 
Figure ~\ref{fig:quantized_contrast} follows unquantized weight distribution plots from Figure ~\ref{fig:unquantized_shift} and plots these blue pillars on the background of each subplot. The orange pillars in the foreground signify the quantized weight distribution shift. The first row is for the distribution shift of the INT4 quantization (with four sampled sections: before training, after the second, the third, and the fifth epoch), while the second row is for the distribution shift of the INT8 quantization.} 
\label{fig:quantized_contrast} 
\end{figure*} 

\begin{figure*} 
\includegraphics[width=\textwidth]{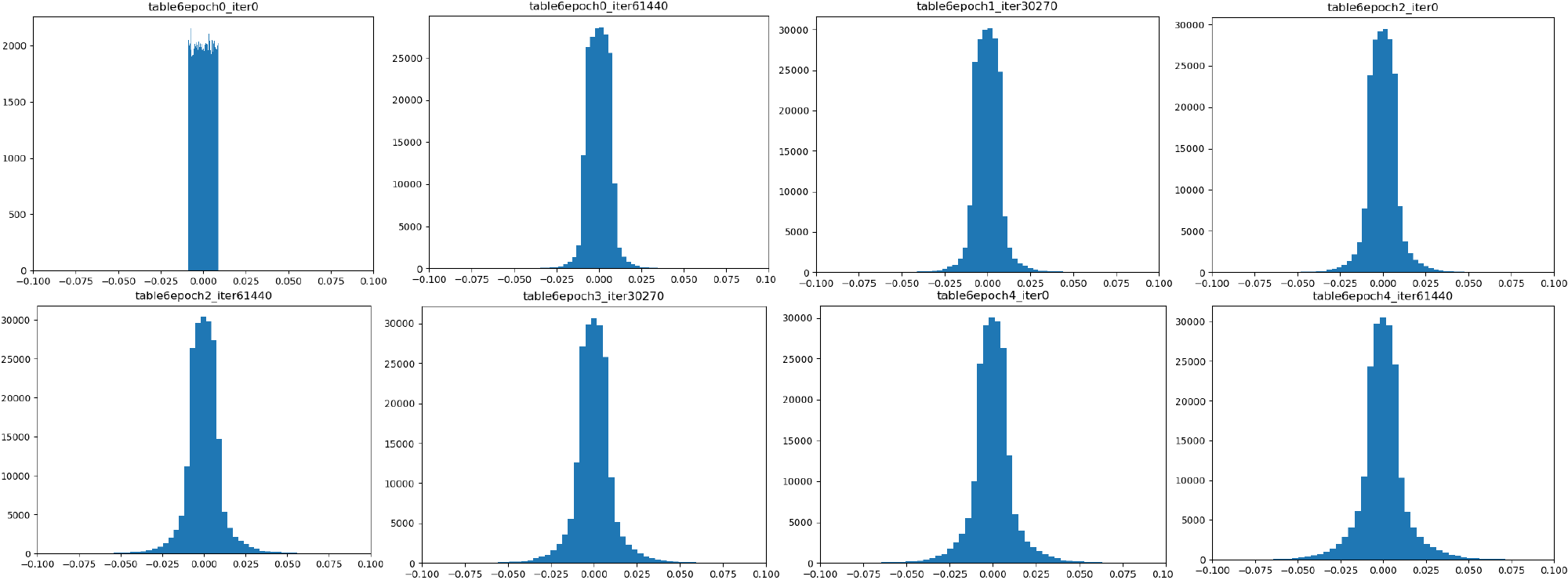} 
\caption{The unquantized weight distribution shift across five epochs. The unquantized model starts overfit after the first epoch}
\label{fig:unquantized_shift} 
\end{figure*} 

To provide more intuition of why INT4 quantization can provide stronger resilience toward overfitting. We look into the weight distribution shift of normal unquantized training, the QAT in INT8 training, and the DQRM in INT4. 
We select Table 6 among the 26 tables from the DLRM Kaggle setting because Table 6 is a medium size embedding table. We found that large tables usually have the majority of weights distributed around 0. However, we found that the medium size embedding tables have a similar trend to large embedding tables. The model gradually shifted outwards from a uniform distribution probability mass function into a bell-shaped curve. The large embedding table has a trend similar to that, but still, after a few epochs, the majority of weight still gathers around 0. In comparison, mid-sized embedding tables display a more obvious weight distribution shift, so we select table 6, a mid-sized embedding table among 26 tables. 

Across the 5 epochs, we sampled four sections in time: before training, at the end of the second, the third, and the fifth epoch. These four sections are aligned on the row. The contrast between rows is for the contrast between different quantization bit-widths of the Quantization-aware training. 
The weight gradually moves from a uniform initialization into a bell-shaped curve. Figure ~\ref{fig:quantized_contrast} contrasts quantized weights (orange, foreground) and unquantized weights (blue background) distribution. INT8 traces the unquantized weight distribution much closer than INT4 quantization.

The general trend of weight distribution shift during normal DLRM training is to gradually shift outward, which can be noted by the contrast of the blue histogram between (b) the weight distribution after the second epoch and (d) after 5 epochs. We observed that the quantized weights (orange) exhibit a similar outward-shifting trend. DQRM INT4 quantized weights (orange) in (b) and (d) seemingly trace the unquantized weight poorly compared with INT8 weights in (c), having the majority of the weight distributed near the origin while having some portion of the weights located symmetrically around 0. In comparison, (c) shows INT8 can roughly cover the entire weight distribution neatly. We also observe that the FP32 copy of weights under the quantization-aware training regardless of bitwidth are highly similar in distribution compared with normal training. 
As shown in section ~\ref{quant_emb_tables}, INT8 although converges slightly higher than the unquantized model, still follows a similar overfitting trend. We hypothesized that the INT4 quantization resilience towards overfitting comes from its large alienation to normal embedding table weight distribution shift. Also, we observe that the INT4 model has the majority of weights close to zero while performing well, which provokes the thought of whether high sparsity in large embedding tables exists and can be better utilized to tackle overfitting. We leave the question to explore later in the work. 

\subsection{Training DQRM for a Large Number of Epochs} 
\label{longepochs} 

\begin{figure}[ht] 
\centering 
\includegraphics[width=\textwidth]{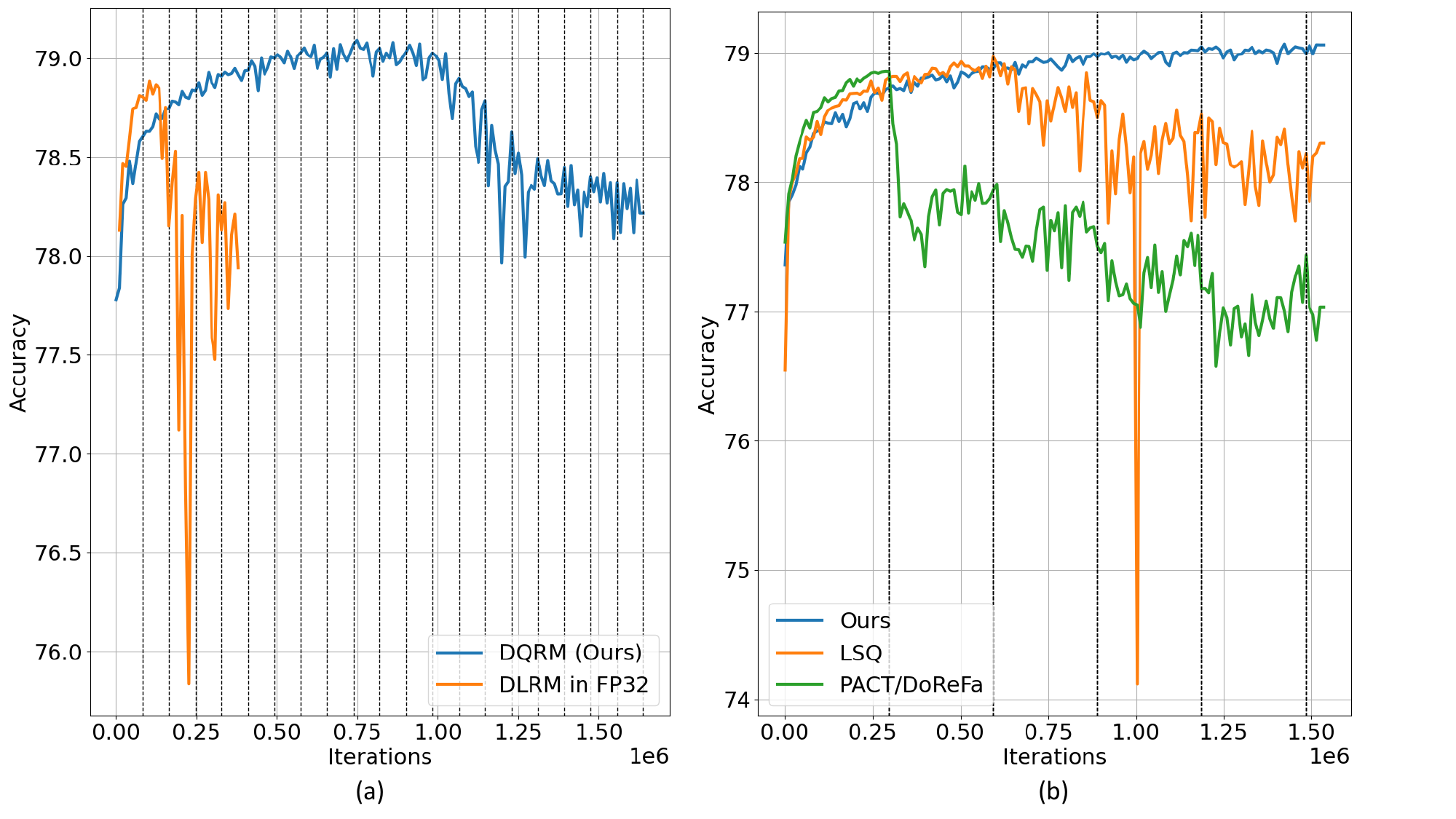} 
\caption{Figure (a) shows the observations when running DQRM INT (blue) for 20 epochs versus running DLRM in FP32 for 5 epochs; Figure (b) compares DQRM, LSQ, and PACT;} 
\label{20epochsfigure} 
\end{figure} 

In our main body, we observed that DQRM INT4 prevents overfitting through five training epochs. For reference, the baseline DLRM in FP32 overfits after one or two epochs. To what extent will DQRM INT4 prevent overfits? Are five epochs enough? To what extent the testing accuracy will plateau? Motivated by these questions, we run DQRM INT4 for an extended 20 epochs till it eventually overfits. We present the testing accuracy curve in Figure ~\ref{20epochsfigure} (a) contrasting DLRM in FP32 (orange curve) and DQRM INT4 (blue curve). Below we also summarize the specific testing performance in Table ~\ref{tab:test_long_epochs_training}. Training convergence reaches its peak at the 11th epoch. After that, it plateaus for 2 to 3 epochs and then goes down. Our experiment shows that 5 epochs aren't enough to make DQRM INT4 fully converged. However, still, considering spending twice the amount of time for a slight accuracy advantage, we leave our results here for the user to decide which scenario they are willing to go for. 

\begin{table}[htbp]
  \centering
  \caption{Test results by number of epochs}
    \begin{tabular}{ccc}
    \toprule 
    \textbf{\# Epochs} & \textbf{Test Accuracy} & \textbf{Test ROC AUC} \\
    \midrule 
    1     & 78.581\% & 0.7967 \\ 
    5     & 78.949\% & 0.8050 \\ 
    10    & 79.092\% & 0.8073 \\ 
    \hc Peak (epoch 11)  & 79.092\% & 0.8076 \\ 
    15    & 78.743\% & 0.8009 \\ 
    20    & 78.219\% & 0.7499 \\ 
    \bottomrule 
    \end{tabular}
  \label{tab:test_long_epochs_training} 
\end{table} 

\subsection{Prior QAT Techniques on DLRM}

In this section, we try to address why different prior QAT methods underachieved in the DLRM model settings. In the main body, we also compare DQRM against HAWQ. Since DQRM without all the tricks added additionally was built on top of the mere quantization scheme of HAWQ. Therefore, HAWQ directly faces the overwhelmingly significant memory usage and long training time for the Terabyte dataset to converge. What we want to discuss here is more focused on other QAT techniques such as PACT and LSQ. Quantization is key in building a mapping between quantized integer values and their original real values. PACT introduces the learned clipping range during activation quantization, while the weight quantization is completely reliant on DoReFa quantization. DoReFa uses the tanh function extensively, but empirically the mapping is not comparable to uniform mapping implemented by the vast majority of quantization works that follow, such as LSQ or HAWQ. Shown in Figure ~\ref{20epochsfigure} (b), it converges the fastest among all techniques and suffers from overfitting severely. LSQ on the other hand learns its clipping range during weight quantization. However, it is not as competitive as DQRM which uses static min max clipping range on both datasets. A potential explanation is that in \ref{dist_shift} of the supplemental materials, we present the weight distribution shift of the DLRM model during training. The range of values is spreading out constantly throughout, unfortunately for LSQ, the optimal clipping range might happen to be a moving objective, which hurts its performance. DQRM's static min-max clipping range might benefit from its stronger ability to adapt. 
\label{priorqat}